\documentclass[twocolumn,prl,aps,floatfix,superscriptaddress]{revtex4}
\usepackage[latin1]{inputenc}
\usepackage{graphicx}
\usepackage{amssymb}
\usepackage{amsmath}
\usepackage{xspace}
\usepackage{bm}
\usepackage{times}
\usepackage{color}
\usepackage{fancyhdr}

\begin{document}

\unitlength=1mm

\title{Non-equilibrium delocalization-localization transition of photons in circuit QED}

\author{S.\ Schmidt}
\affiliation{Institute for Theoretical Physics, ETH-Zurich, CH-8093 Zurich, Switzerland}
\author{D.\ Gerace}
\affiliation{Dipartimento di Fisica ``Alessandro Volta'', Universit\`a di Pavia, 27100 Pavia, Italy}
\author{A.\ A.\ Houck}
\affiliation{Department of Electrical Engineering, Princeton University, Princeton, NJ 08544, USA}
\author{G.\ Blatter}
\affiliation{Institute for Theoretical Physics, ETH-Zurich, CH-8093 Zurich, Switzerland}
\author{H.\ E.\ T\"ureci}
\affiliation{Department of Electrical Engineering, Princeton University, Princeton, NJ 08544, USA}
\affiliation{Institute for Quantum Electronics, ETH-Zurich, CH-8093 Zurich, Switzerland}

\date{\today}

\begin{abstract}
We show that photons in two tunnel-coupled microwave resonators each containing a single superconducting qubit undergo 
a sharp non-equilibrium delocalization-localization (self-trapping) transition due to strong photon-qubit coupling. 
We find, that self-trapping of photons in one of the resonators (spatial localization) 
forces the qubit in the opposite resonator to remain in its initial state (energetic localization).
This allows for an easy experimental observation of the transition by local read-out of the qubit state.
Dissipation of photons and decoherence of the qubit favor the self-trapped regime.
\end{abstract}


\maketitle

In circuit quantum electrodynamics (QED), superconducting qubits
are coupled with microwave photons in a transmission line resonator 
reaching extremely strong light-matter interactions within an integrated circuit \cite{WS04}. 
Device integration, high tunability and individual addressability of each resonator make wide parameter 
regimes easily accessible. Circuit QED thus constitutes 
one of the most promising solid-state architectures for quantum information processing 
and offers the possibility to study fundamental questions of interacting quantum systems \cite{SG08}.  
The experimental focus has been on the design of the coupling between a single cavity and a single qubit
and subsequent work demonstrated a great level of control of single-cavity systems \cite{SH07,SP07,HW08,BC09,JR10}.
Today, a key challenge for scalability and further progress in the field is the understanding of 
small coupled systems, i.e., effective qubit-qubit and photon-photon interactions and
their interplay with dissipation \cite{MC07,FB09,DG09}. 

Recent theoretical interest in hybrid light-matter systems has been on the superfluid-Mott insulator transition of polaritons
in a coupled-cavity array \cite{GT06,HB06,AS07,RF07,AH08,SB09,KH09,SB10}.
Although the theoretical investigation of this quantum phase transition has triggered enormous interest, the experimental realization of a large array of identical cavities is very ambitious. In this letter we show that even in the smallest possible coupled-cavity system of a Jaynes-Cummings dimer (JCD), one can find strong signatures of the on-site repulsive interaction among photons. 
The two-coupled cavity system proposed here is simple enough to be readily realizable with state-of-the-art circuit QED technology.

\begin{figure}[b]
\includegraphics[width=0.35\textwidth,clip]{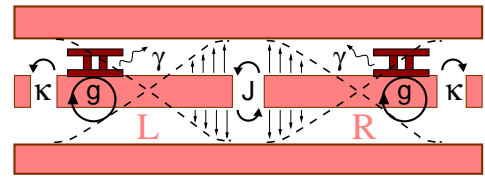}
\vspace{-0cm}
\caption{\label{fig1} Schematic diagram of the photonic Josephson junction (PJJ) proposed in this paper. Two transmission line microwave resonators (L,R) are coupled in series with a tunneling rate $J$, determined by the series capacitance of the resonators.  Each resonator is strongly coupled to a superconducting qubit with a coupling rate $g$, providing a strong JC non-linearity. Photons can leave each resonator at a rate $\kappa$, providing a mechanism for dissipation. Decoherence of the qubit is taken into account via the spontaneous emission rate $\gamma$.}
\vspace{-0cm}
\end{figure}

We study a photon Josephson junction (PJJ) consisting of two
tunnel-coupled microwave resonators each containing a single superconducting qubit (Fig.~\ref{fig1}).
We show that photons undergo a sharp non-equilibrium delocalization-localization
transition from a regime where an initial photon population imbalance between the two resonators 
undergoes coherent oscillations (delocalized) between the 
two resonators to another regime where it becomes
self-trapped (localized) as the photon-qubit interaction is increased.
Similar self-trapping transitions were found in optical fibres \cite{Je82}, molecules \cite{EL85},
cold atom \cite{AS97,MA05,SL07} and polariton BEC's \cite{SC08}. In all of these systems self-trapping 
is due to a Kerr/Bose-Hubbard like nonlinearity and has been experimentally observed in the semiclassical regime with a large number of particles. 
The circuit QED implementation proposed in this paper has several advantages with respect to these systems:
(i) Self-trapping is due to a Jaynes-Cummings (JC) rather than a Kerr/Bose-Hubbard like nonlinearity. The JC 
interaction accurately describes the photon-qubit coupling in a microwave resonator \cite{JF08}. 
In contrast, a Kerr/Bose-Hubbard like nonlinearity is often a rather crude approximation of the experimental conditions \cite{ZS03,MP09} 
(ii) The PJJ is a genuinely dissipative system. We show that dissipation and spontaneous emission of the qubit favor the localized regime.
(iii) The PJJ may operate in the semiclassical (many photons) as well as quantum (few photons) limit since each resonator can initially be pumped with an 
almost arbitrary number of photons.\\
In the following, we study in detail the classical versus quantum nature of this transition and present numerical as well as analytical results
including the effects of dissipation and decoherence. At the end of the paper, we outline a precise proposal on how to measure the localization transition
of photons experimentally.

We describe the PJJ (Fig.~\ref{fig1}) by a two-site Jaynes-Cummings-Hubbard Hamiltonian (JCHM) 
\begin{equation}
\label{jcd}
H=\sum_{i={\rm L,R}}  h^{\rm JC}_i - J (a^\dagger_{\rm L} a_{\rm R} + {\rm h.c.})\,,
\vspace{-0.1cm}
\end{equation}
where $h^{\rm JC}_i$ denotes the local JC Hamiltonian 
$h^{\rm JC}_i = \omega_c\, a^\dagger_i a_i + \omega_x \sigma_{i}^+\sigma_{i}^-  +  g (\sigma_{i}^+ a_i +\sigma_{i}^- a^\dagger_i)$
for the left $({\rm L})$ or the right (${\rm R}$) cavity, $a_i$ ($a_i^\dagger$) and $\sigma_i^{+}$ ($\sigma_i^{-}$) are the photon creation (annihilation) and qubit raising (lowering) operators, respectively. 
The photon mode frequency is $\omega_c$, the qubit transition energy is $\omega_x$ and the photon-qubit coupling is given by $g$ (we set $\hbar=1$).  
The photon-qubit interaction induces an anharmonicity in the spectrum of the JC Hamiltonian which leads to an effective on-site repulsion (anti-bunching)
for photons. Throughout the paper we will assume zero detuning ($\omega_x=\omega_c$) for which this anharmonicity is strongest.
Recently, dynamical aspects in a finite size JCHM have  been investigated, but restricted to the one excitation (either photon or qubit) subspace (i.e., photon-photon interaction effects are irrelevant) and neglecting dissipation \cite{OI08,MC09,ZL10}. 
\begin{figure}
\includegraphics[width=0.5\textwidth,clip]{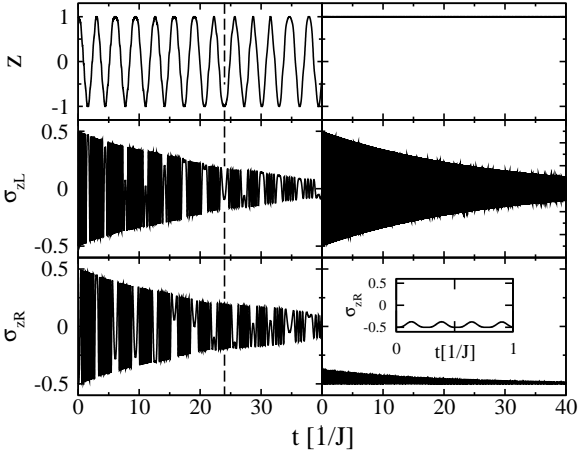}
\vspace{-0.9cm}
\caption{\label{fig2} Photon imbalance $z(t)$ and inversion of the qubits $\sigma_{z\rm (L,R)}(t)$ obtained from a semiclassical approximation for the dissipative JCD with 
$\kappa=\gamma=0.05 J$. Shown are results in the Josephson regime with $g=0.4 g_c$ (left)  and in the self-trapped regime with $g=2g_c$ (right). 
The vertical dashed line points to a situation where all photons have tunneled from the left to the right cavity and thus Rabi oscillations of the left qubit have slowed down while the right
qubit oscillates rapidly. The inset in the right panel shows the inversion of the right qubit at short times in the self-trapped regime displaying strongly suppressed Rabi oscillations. Here, we have chosen the initial condition $\psi_L(0)=\sqrt{20}$ corresponding to a coherent initial state with average photon number $N=20$.}
\vspace{-0.4cm}
\end{figure}
\begin{figure}
\includegraphics[width=0.45\textwidth,clip]{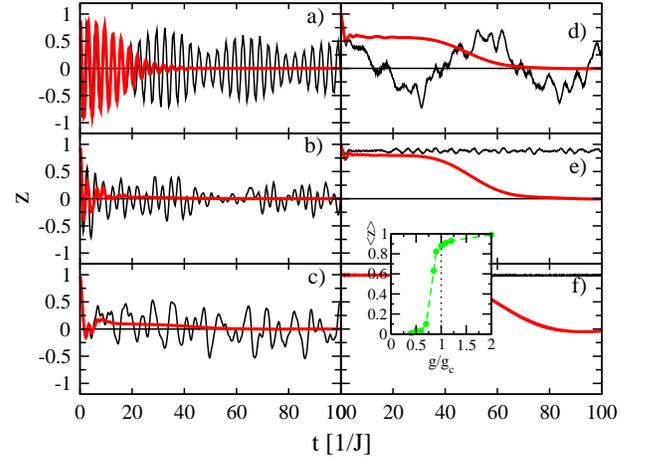}
\vspace{-0.2cm}
\caption{\label{fig3} Photon imbalance $z(t)$ obtained from a full numerical solution of the master equation (\ref{master}) (including quantum fluctuations) with $\kappa=\gamma=0$ (black thin curves) 
and $\kappa=\gamma=0.05 J$ (red thick curves) for initially $N(0)=20$ photons (and qubits initially in their groundstate $\sigma_{z\rm(L,R)}(0)=-1/2$).
Figs.~(a)-(f) show results for $g=0.1,0.4,0.6,0.8,0.9,2g_c$. The inset shows the time-averaged imbalance $\langle z \rangle$ (averaged over the time interval $t \in [0,100/J]$) for $\kappa=\gamma=0$) (dashed line) as a function of the photon-qubit coupling $g$ normalized with the semiclassical critical value $g_c$ in Eq.~(\ref{gcrit}). In comparison, the semiclassical transition at this scale is essentially abrupt (dotted line).}
\vspace{-0.4cm}
\end{figure}

In the JCD discussed here, photon dissipation and qubit decay are both taken into account by a Lindblad master equation for the system's density matrix $\rho$
\begin{eqnarray}
\label{master}
\frac{\partial \rho}{\partial t} = i [\rho, H] + \sum_{i={\rm L,R}} ( \kappa\mathcal{L}[a_i] + \gamma\mathcal{L}[\sigma^-_i] )\,,
\end{eqnarray}
where the Liouvillian of an operator $O$ is defined as $\mathcal{L}[O] =\left( 2 O \rho O^\dagger - O^\dagger O \rho - \rho O^\dagger O\right)/2$.
Here, $\kappa$ and $\gamma$ denote cavity decay and spontaneous photon emission rate by the qubit, respectively, and we neglected pure qubit dephasing as it can be experimentally suppressed with respect to the other two decay channels, e.g. in a transmon qubit \cite{KY07}. Both decay processes in (\ref{master}) lead to a decrease of the total number of photons as a function of time.

The central quantity of interest is the photon population imbalance $z(t)=(n_{\rm L}(t)-n_{\rm R}(t))/N(t)$ with $n_i=\rm{Tr}\,\hat{a}^\dagger_i\hat{a}_i \hat{\rho}$ and the total photon number $N=n_{\rm L}+n_{\rm R}$. 
Throughout the paper, we consider the experimentally most relevant initial condition where the left cavity is pumped with $N$ photons, the right cavity is empty and both qubits reside in their respective ground-state at $t=0$.
For large photon numbers, we can resort to a semiclassical approximation in which correlation functions in Eq.~(\ref{master}) are decoupled
by simple factorization, e.g., $\langle a^\dagger \sigma^-\rangle \approx \langle a^\dagger \rangle \langle \sigma^-\rangle$, yielding eight
coupled equations of motion for the expectation values of the photon and qubit operators which can be solved for an arbitrary
number of photons \cite{FN}. In the following, we present results of the semiclassical approximation (Fig.~\ref{fig2}) as well as the full quantum solution of Eq.~(\ref{master}) (Fig.~\ref{fig3}).

We first discuss results of the semiclassical approximation without dissipation ($\kappa=\gamma=0$). 
In this case one can further reduce the number of coupled equations using energy
conservation and the specific initial condition chosen above, yielding only four equations of motions
\begin{eqnarray}
\label{semicl}
\dot{\theta}_{\rm L} &=& -2 g {\rm Re}(\psi_{\rm L})\nonumber\\
\dot{\theta}_{\rm R} &=& -2 g {\rm Im}(\psi_{\rm R})\nonumber\\
{\rm Re}(\dot{\psi}_{\rm L}) &=& g\sin(\theta_{\rm L})/2 - J {\rm Im}(\psi_{\rm R})\nonumber\\
{\rm Im}(\dot{\psi}_{\rm R})&=& g\sin(\theta_{\rm R})/2 + J {\rm Re}(\psi_{\rm L})\,,
\end{eqnarray}
where $\psi_{\rm(L,R)}=\langle a_{\rm(L,R)}\rangle $, ${\rm Im}(\psi_{\rm L})={\rm Re}(\psi_{\rm R})=0$ and the angles $\theta_i$ describe the qubit states $\langle\vec{\sigma}_{\rm L}\rangle=-(\sin \theta_{\rm L}, 0, \cos \theta_{\rm L})/2$ and $\langle\vec{\sigma}_{\rm R}\rangle=-(0, \sin \theta_{\rm R}, \cos \theta_{\rm R})/2$.
At zero interaction ($g=0$) Eqs.~(\ref{semicl}) are exactly solvable. In this limit the photon imbalance
undergoes coherent harmonic oscillations $z(t)=\cos \omega_J t$ with frequency $\omega_J=2J$. 
As $g$ becomes non-zero this frequency decreases and the oscillations become anharmonic. 
At a critical value of the coupling constant
\begin{eqnarray}
\label{gcrit}
g_c\approx 2.8 \sqrt{N} J
\end{eqnarray} 
the period of oscillation diverges (critical slowing down) and an abrupt transition
occurs to a localized regime, where the initial photon imbalance stays almost completely in the left cavity, i.e., $z(t)\approx 1$. 
Solutions of Eq.~(\ref{semicl}) near the transition ($g\sim g_c$) are shown in Fig.~\ref{fig2}. 
An important and useful result is that the localization transition of photons can be observed in the population inversion of the two qubits, which depend on the number of photons in each cavity. In the delocalized regime ($g<g_c$, see Fig.~\ref{fig2} left), when photons are tunneling, e.g., from the left into the right cavity, Rabi oscillations of the left qubit slow down considerably while those of the right qubit speed up. After half a tunneling period the scenario is reversed. On the other hand, in the localized regime ($g>g_c$, see Fig.~\ref{fig2} right) the left qubit displays fast, complete Rabi oscillations while the right one displays slow, small amplitude oscillations (right column in Fig.~\ref{fig2}). In other words, spatial localization of photons in the left resonator induces an energy state localization of the right qubit (notice that deep in the localized regime ($g\gg g_c$) the right qubit remains very close to the ground-state at all times, i.e., $\sigma_{z\rm(L,R)}(t)\approx\sigma_{z\rm(L,R)}(0)$). This suggests that the localization transition of photons in a PJJ can be observed experimentally by a local readout of the qubit states. It also shows that qubit-qubit correlations are largely suppressed in the localized regime. 

Taking into account dissipation and spontaneous emission within the semiclassical approximation (i.e., solving all eight coupled equations of motion) leads to mainly three effects:
(i) The oscillations of the qubit inversion are damped due to spontaneous emission. 
(ii) The frequency of these oscillations slow down with time as photons are leaving the cavity.
(iii) Dissipation and spontaneous emission make the interaction terms of the Hamiltonian effectively time dependent. If the parameters of the system are properly chosen this can lead to a 
transition from a delocalized to a localized phase at finite times. Thus, dissipation and spontaneous emission, stabilize the localized regime, i.e., we find self-trapped solutions for significantly 
smaller values of the photon-qubit coupling $g<g_c$. In Fig.~\ref{fig2} we have chosen the coupling strengths such that we always stay within one phase; a careful analysis of this switching behavior will be given elsewhere.
\begin{figure}
\begin{minipage}[t]{4.8cm}
\centering
\includegraphics[width=4.8cm,clip]{fig4ab.eps}
\end{minipage}
\hfill
\begin{minipage}[t]{3.5cm}
\centering
\includegraphics[width=3.1cm]{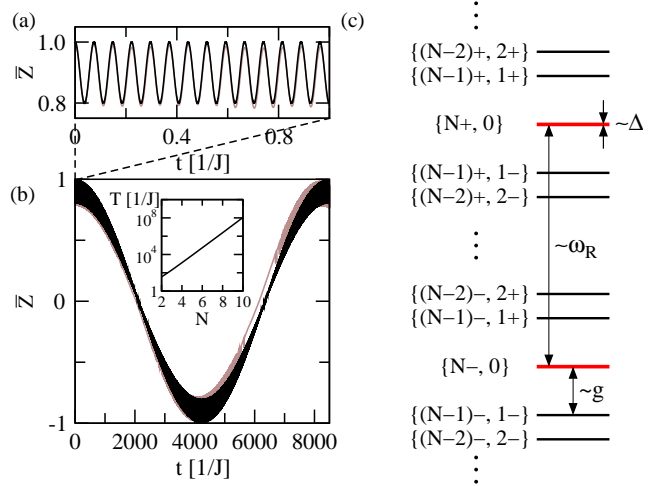}
\end{minipage}
\caption{\label{fig4} Rescaled photon imbalance $\tilde{z}(t)=(n_1(t)-n_2(t))/N(0)$ deep in the self-trapped regime for $N(0)=5$ photons
with $g=3g_c$ and $\kappa=\gamma=0$. Shown are small amplitude Rabi oscillations on short time scales (Fig.~\ref{fig4}(a)) and large amplitude ultra-long tunneling (Fig.~\ref{fig4}(b)). 
Results of a full numerical solution of the quantum master equation (\ref{master}) (grey curve in Figs.~\ref{fig4}(a) and \ref{fig4}(b))
are compared with strong-coupling degenerate perturbation theory (sdPT) (black curve) based on the effective level scheme shown in in Fig.~\ref{fig4}(c). 
Note that a polariton eigenstate $|M\sigma\rangle$ is a mixed photon ($M,M-1$) - qubit ($g,e$) state, i.e, $|M\sigma\rangle=(|M\,,g\rangle +\sigma |(M-1)\,,e\rangle)/\sqrt{2}$ 
(the zero polariton state is a special case with $|0\rangle\equiv|0-\rangle=|0\,,g\rangle$).
In Fig.~\ref{fig4}(c), ${M\sigma,K\mu}=\{|M\sigma\rangle_L |K\mu\rangle_R, |K\mu\rangle_L |M\sigma\rangle_R\}$ denotes the pair of degenerate 
polariton eigenstates of the Hamiltonian (\ref{jcd}) at $J=0$ with ($M,K$) lower/upper ($\sigma,\mu=\pm$) polaritons in left (L) and right (R) cavity, respectively. 
Their degeneracy is lifted due to tunneling $J$, which induces the splitting $\Delta$ leading to ultra-long tunneling with period $T=2\pi/\Delta$ (see inset in Fig.~\ref{fig4}(b)).}
\vspace{-0.5cm}
\end{figure}

Full numerical solutions (including quantum fluctuations) of the master equation (\ref{master}) using a Fock state basis for $20$ photons are shown in Fig.~\ref{fig3}. 
Most importantly, we observe that the localization transition survives. However, specific quantum correction show up:
(i) Deep inside the delocalized regime (see Fig.~\ref{fig3}(a)) the photon imbalance displays beatings of the coherent oscillations,
a quantum feature that is absent in semiclassical solutions. Dissipation and spontaneous emission strongly damp the large amplitude oscillations and suppress these beats;
(ii) the localization transition is shifted to smaller $g$ values and smoothened (see inset Fig.~\ref{fig3});
(iii) in the localized regime with dissipation and spontaneous emission (see Figs.~\ref{fig3}(e,f)) the imbalance approaches zero asymptotically at long times displaying no single-particle
Rabi oscillations (i.e., when all photons left both cavities). 
(iv) the deeply localized regime displays rich multi-scale time dynamics, which can be explained using the effective level scheme shown in Fig.~\ref{fig4}(c). 
For short times, Rabi oscillations of the qubit induce small amplitude oscillations (on the order $\sim 1/N$) of the rescaled photon imbalance $\bar{z}(t)=(n_1(t)-n_2(t))/N(0)$, see Fig.~\ref{fig4}(a). 
The frequency of the Rabi oscillations (due to exchange of one photon between local qubit and cavity) is given by the large splitting between lower and upper polariton states which yields for
$J=0$ and large photon numbers $\omega_R=2g\sqrt{N}+\mathcal{O}(1/N)$. For ultra-long times the localization of photons is unstable and almost complete oscillations of the imbalance set in, see Fig.~\ref{fig4}(b). 
Neglecting the perturbative corrections to the eigenvectors of the Hamiltonian (\ref{jcd}), we find for the rescaled imbalance
$\bar{z}=\cos(\Delta t) \left[ 1-(1/N)\sin^2(\omega_R t)\right]+\mathcal{O}(J^2/g^2)$ which quantitatively reproduces the numerical results in Figs.~4(a) and 4(b) (red curve).
The period of ultra-long tunneling is set by the splitting $\Delta$, which we have calculated
in leading order from $N$-th order degenerate strong-coupling perturbation theory, yielding 
$\Delta= c_N J (J/g)^{N-1}$ with a constant $c_N$ that depends on the number of photons $N$. The inset in Fig.~4(b) shows the scaling of the corresponding time period $T=2\pi/\Delta$ of the large amplitude oscillations
as a function of $N$. The period $T$ increases very fast (quasi-exponentially) with increasing photon number (see inset in Fig.~\ref{fig4}(b)).
Thus, true localization disappears in a small system, but turns exponentially 'good' with increasing system size (photon number).
Already for five photons the ultra-long tunneling regime is hardly accessible experimentally. Ultra-long tunneling times were also predicted for 
the BHM \cite{BE90}. We should note, however, that under experimental conditions any asymmetry between the two cavities (e.g., due to small differences in detunings
or coupling constants) will lift this degeneracy and set the effective time period of the large amplitude oscillations. 
A corollary of this statement is that as long as such an asymmetry is much smaller
than the frequency $\omega_J=2J$ the localization transition should be observable.

The physics of a PJJ should be readily observable using a circuit QED implementation with realistic device parameters.  A possible device consists of two series-coupled transmission line resonantors, each containing a single superconducting qubit (Fig.~\ref{fig1}). A broad parameter space is available through changes in lithographic patterning. In particular, qubit-cavity coupling $g$ can range from 1 MHz to 300 MHz, while the cavity-cavity coupling $J$ and the cavity dissipation rate $\kappa$ can be tuned independently in a range 50 kHz to 50 MHz. The spontaneous emission rate $\gamma$ is typically 50KHz to 500kHz.
An experimental observation of the localization transition proceeds in three parts.  One cavity is populated with photons (initialization), the evolution proceeds for a fixed duration of time (evolution), and the photon occupancy of each cavity is finally measured (read-out). 
This entire process would be repeated for varying evolution times, thus allowing full reconstruction of the population imbalance $z(t)$. Initialization can be accomplished using three different methods: 
(i) In the simplest method, the cavity is populated with a coherent photon state while the qubit is far off resonance ($\omega_x\ll\omega_c$); the qubit is then quickly brought into resonance ($\omega_x=\omega_c$) for the evolution \cite{SP07, HW08}. This scenario is best described by the results of the semiclassical approximation in Fig.~\ref{fig2}.
(ii) The cavity-qubit system can also be populated directly with a few-polariton state, i.e., an eigenstate of the JC Hamiltonian, using a properly timed $\pi$-pulse \cite{BC09}.
(iii) Finally, a $N$-photon Fock state can be constructed sequentially by successively exciting the qubit very quickly and bringing it into resonance ($\omega_x=\omega_c$).
The multi-photon/polariton transitions in the latter two scenarios are resolvable up to $5-10$ excitations and are faithfully realized by the initial conditions chosen in Fig.~\ref{fig3} (full quantum calculation). 
After evolution for a given time, the qubits coupled to each cavity will be used to measure the photon occupation of that cavity. When strongly coupled, the qubit frequency is shifted depending on the number of photons in the cavity \cite{SH07}; by interrogating these different frequencies, a quantum non-demolition experiment of the photon number can be performed \cite{JR10}.  

In this work we have shown that two tunnel-coupled microwave resonators each containing a Jaynes-Cummings type nonlinearity
undergo simultaneously sharp localization transitions of photons (spatial) and qubits (energetic).  Our results suggest many directions of further theoretical investigations
including effects of detuning, quantum-classical crossover and interplay of localization and entanglement.

\begin{acknowledgments}
We thank S. Girvin and J. Koch for discussions and acknowledge financial support from the Swiss National Foundation through the NCCR MaNEP and Grant No. PP00P2-123519/1. 
\end{acknowledgments}

\vfill

\begin{thebibliography}{99}
\bibitem{WS04} A. Wallraff {\it et al.}, Nature {\bf 431}, 162 (2004).
\bibitem{SG08} R. J. Schoelkopf and S. M. Girvin, Nature {\bf 451}, 664 (2008).
\bibitem{SH07} D. I. Schuster {\it et al}, Nature 445, 515-518 (2007).
\bibitem{SP07} M. A. Sillanpaa {\it et al}, Nature 449, 438-442 (2007).
\bibitem{HW08} M. Hofheinz {\it et al}, Nature 454, 310-314 (2008).
\bibitem{BC09} L. S. Bishop {\it et al}, Nat. Phys. 5, 105-109 (2009).
\bibitem{JR10} B. R. Johnson {\it et al}, arxiv:1003.2734 (2010).
\bibitem{JF08} J. M. Fink {\it et al}, Nature {\bf 454}, 315 (2008).
\bibitem{MC07} J. Majer {\it et al.}, Nature {\bf 449}, 443 (2007).
\bibitem{FB09} J. M. Fink {\it et al}, Phys. Rev. Lett. {\bf 103}, 083601 (2009).
\bibitem{DG09} D. Gerace {\it et al}, Nature Phys. {\bf 5}, 281 (2009). 
\bibitem{GT06} A. D. Greentree {\it et al.}, Nature Phys. {\bf 466}, 856 (2006).
\bibitem{HB06} M. J. Hartmann, F. G. S. L. Brandao, and M. B. Plenio, Nature Phys. {\bf 462}, 849 (2006).
\bibitem{AS07} D. G. Angelakis, M. F. Santos, and S. Bose, Phys. Rev. A {\bf 76}, 031805(R) (2007).
\bibitem{RF07} D. Rossini and R. Fazio, Phys. Rev. Lett. {\bf 99}, 186401-1 (2007).
\bibitem{AH08} M. Aichhorn, M. Hohenadler, C. Tahan, and P. B. Littlewood, Phys. Rev. Lett. {\bf 100}, 216401 (2008).
\bibitem{SB09} S. Schmidt and G. Blatter, Phys. Rev. Lett. {\bf 103}, 086403 (2009).
\bibitem{KH09} J. Koch and K. Le Hur, Phys. Rev. A {\bf 80}, 023811 (2009).
\bibitem{SB10} S. Schmidt and G. Blatter, Phys. Rev. Lett. {\bf 104}, 216402 (2010).
\bibitem{Je82} S. M. Jensen, IEEE J. Quantum Electron. QE-18, 1580 (1981)
\bibitem{EL85} J.C. Eilbeck, P.S. Lomdahl and A.C. Scott, Physica D {\bf 16}, 318 (1985).
\bibitem{AS97} A. Smerzi {\it et al.}, Phys. Rev. Lett. {\bf 79}, 4950 (1997).
\bibitem{MA05} M. Albiez {\it et al.}, Phys. Rev. Lett. {\bf 95}, 010402 (2005).
\bibitem{SL07} S. Levy {\it et al.}, Nature Lett. {\bf 449}, 579 (2007).
\bibitem{SC08} D. Sarchi {\it et al}, Phys. Rev. B {\bf 77}, 125324 (2008).
\bibitem{ZS03} I. Zapata, F. Sols, and A. J. Leggett, Phys. Rev. A {\bf 67}, 021603(R) (2003).
\bibitem{MP09} M. T. Martinez, A. Posazhennikova, and J. Kroha, Phys. Rev. Lett. {\bf 103}, 105302 (2009).
\bibitem{OI08} C. D. Ogden, E. K. Irish, and M. S. Kim, Phys. Rev. A {\bf 78}, 063805 (2008).
\bibitem{MC09} M. I. Makin {\it et al}, Phys. Rev. A {\bf 80}, 043842 (2009).
\bibitem{ZL10} Ke Zhang and Zhi-Yuan Li, Phys. Rev. A {\bf 81}, 033843 (2010).
\bibitem{BE90} L. Bernstein, J. Eilbeck, A. Scott, Nonlinearity 3, 293. (1990).
\bibitem{KY07} J. Koch {\it et al.}, Phys. Rev. A {\bf 76}, 042319 (2007).
\bibitem{FN} Rescaling of the time parameter $\tilde{t}=t/N$ and the coupling constants $\tilde{g}=g\sqrt{N}$, $\tilde{J}=J N$, $\tilde{\kappa}=\kappa N$ and $\tilde{\gamma}=\gamma N$  
renders the semiclassical equations of motion independent of the initial photon number $N$ since $\psi_L(0)=\sqrt{N}$ and $\psi_R(0)=0$.
\end{thebibliography}
\end{document}